\documentclass[twocolumn,twoside]{revtex4}
\usepackage{graphicx}
\usepackage{fancyhdr}
\def\Journal#1#2#3#4{{#1} {\bf #2}, #3 (#4)}

\begin{document}

\title{Semileptonic and leptonic B decay results from early Belle~II data}

%

\author{Markus Tobias Prim\\
on behalf of the Belle~II Collaboration}
\affiliation{Karlsruhe Institute of Technology (KIT), Institute of Experimental Particle Physics (ETP), Wolfgang-Gaede-Str. 1, 76131 Karlsruhe, Germany}

\begin{abstract}
The Belle II experiment at the SuperKEKB energy-asymmetric $e^+e^-$ collider is a substantial upgrade of the B factory facility at the Japanese KEK laboratory. The design luminosity of the machine is $8\times10^{35}\, \mathrm{cm}^{-2}\mathrm{s}^{-1}$ and the Belle II experiment aims to record $50\, \mathrm{ab}^{-1}$ of data, a factor of 50 more than its predecessor. From February to July 2018, SuperKEKB has completed a commissioning run, achieved a peak luminosity of $5.5 \times 10^{33}\, \mathrm{cm}^{-2}\mathrm{s}^{-1}$, and Belle II recorded a data sample of about $0.5\,  \mathrm{fb}^{-1}$. In this presentation we show first results from studying missing energy signatures, such as leptonic and semileptonic B meson decays based on this early Belle II data. We report first studies on re-measuring important standard candle processes, such as the abundant inclusive $B\rightarrow X l \nu$ and $B\to D^*\ell\nu$ decays, and evaluate the performance of machine learning-based tagging algorithms. Furthermore, we also present an overview of the semileptonic B decays that will be measured in the upcoming years at Belle II and discuss prospects for important B-anomalies like R$(D)$ and R$(D^*)$, as well as other tests of lepton flavor universality.
\end{abstract}

\maketitle

\thispagestyle{fancy}

\section{Introduction}
The Belle~II~\cite{belle2} experiment is located at the SuperKEKB~\cite{superkekb} accelerator in Tsukuba, Japan. The experiment is designed to perform a variety of high-precision measurements, among others, in the field of heavy flavour physics. In particular it is designed to study the decay of $B$ mesons. The Belle~II detector is an upgrade of the Belle detector and is designed to cope with the factor of 40 increased target luminosity of $\mathcal{L}=8 \times 10^{35}\, \mathrm{cm}^{-2}\mathrm{s}^{-1}$ of SuperKEKB.

This manuscript contains three studies on the early Belle~II data from the commissioning run: The performance of the Full Event Interpretation~\cite{fei}, the analysis of inclusive semileptonic $B\rightarrow Xe\nu$ decays, and the analysis of exclusive semileptonic $B\rightarrow D^*e\nu$ decays. Additionally, two prospects of the Belle~II physics program are briefly shown: The projection of the Belle~II sensitivity on $R(D^{(*)})$ and the projected precision of different methods used to extract the CKM matrix element $|V_\mathrm{ub}|$.

\section{Full Event Interpretation Performance}
The Full Event Interpretation (FEI) is a new exclusive tagging algorithm developed for the Belle~II experiment. It allows the measurement of otherwise inaccessible $B$ decays by reconstructing the so-called $B_\mathrm{tag}$ meson. The algorithm reconstructs possible $B_\mathrm{tag}$ decay chains in a hierarchical approach and uses machine learning to identify plausible candidates and to reduce the combinatorics. For a detailed description of the FEI see~\cite{fei}. Here we present the application of the FEI on the Belle~II data sample of $500\, \mathrm{fb}^{-1}$ from the commissioning run. For this study only the hadronic decay chains are considered. 

\begin{figure}
	\includegraphics[width=80mm]{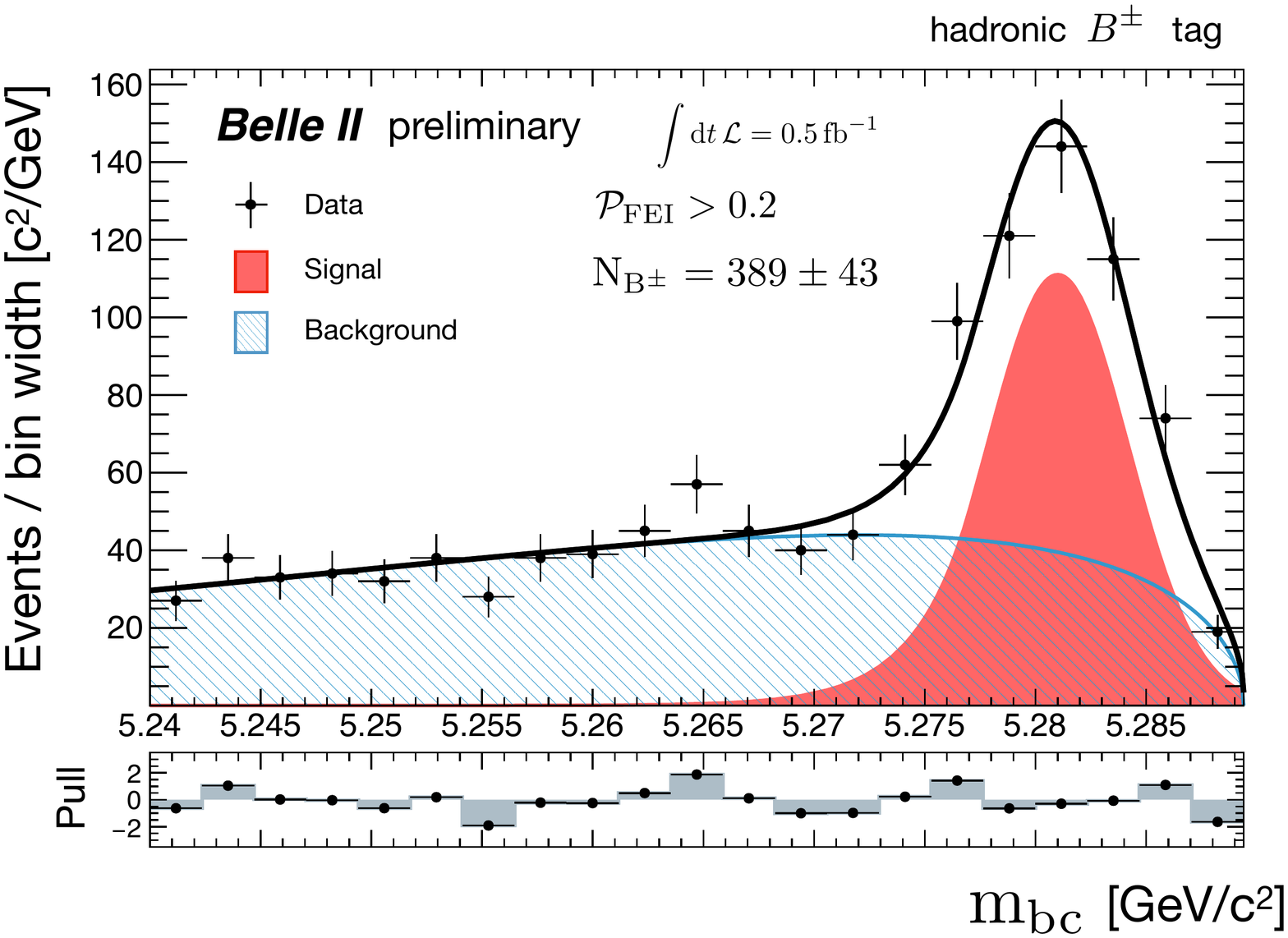}
	\includegraphics[width=80mm]{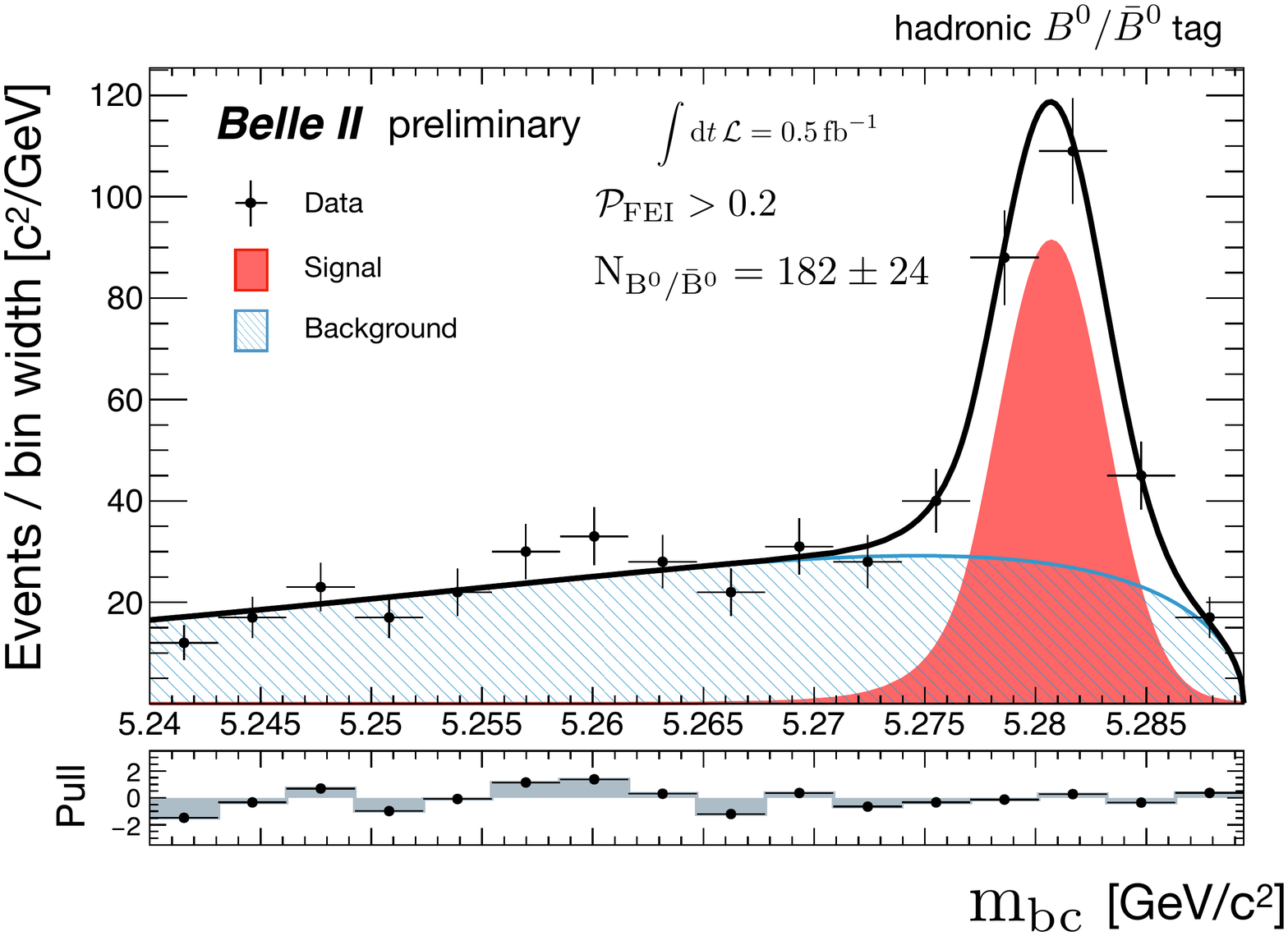}	
	\caption{Charged (top) and neutral (bottom) $B$ meson $m_\mathrm{bc}$ distributions with a tight cut on the signal probability of $\mathcal{P}_\mathrm{FEI} > 0.2$.}
	\label{fig:fei:mbc-fit}
\end{figure}

During the application of the FEI the following cuts on tracks and intermediate particles are applied: Tracks are required to originate from the interaction region $|\mathrm{d}z| < 2\, \mathrm{cm}$, $d_0 < 0.5\, \mathrm{cm}$, and photons are required to have a energy of $E_\gamma > 0.10\, \mathrm{GeV}$ in the forward,  $E_\gamma > 0.09\, \mathrm{GeV}$ in the barrel and  $E_\gamma > 0.16\, \mathrm{GeV}$ in the backward region. An invariant mass cut was performed on the intermediate states $\pi^0$, $D$ and $J/\Psi$ of $0.08\, \mathrm{GeV} < m(\pi^0) < 0.18\, \mathrm{GeV}$, $1.70\, \mathrm{GeV} < m(D) < 1.95\, \mathrm{GeV}$ and $2.8\, \mathrm{GeV} < m(J/\Psi) < 3.5\, \mathrm{GeV}$ respectively. The $D^*$ candidates are accepted if the energy release is $0\, \mathrm{GeV} < Q < 3\, \mathrm{GeV}$. The $B$ candidates need to pass a cut on the beam-constrained mass $m_\mathrm{bc} > 5.22\, \mathrm{GeV}$, the missing energy $0.15\, \mathrm{GeV} < \Delta E < 0.10\, \mathrm{GeV}$, and on the thrust angle $\cos \theta_\mathrm{thrust} < 0.9$. If multiple $B$ candidates survive the selection cuts, the candidate with the highest signal probability is selected. After application of the FEI, the $e^+e^- \rightarrow q\bar{q}$ background was suppressed by a cut on the ratio of the two Fox-Wolfram moments $R_2 = F_2/F_0$, which encodes the topology of the event. To extract the number of reconstructed $B$ mesons a fit was performed on the beam-constrained mass $m_\mathrm{bc}=\sqrt{s/4 - |\vec{p}_\mathrm{tag}^*|^2}$, where $s$ is the center-of-mass energy and $\vec{p}_\mathrm{tag}^*$ the reconstructed momentum of the $B_\mathrm{tag}$ candidate. The fit result for charged and neutral $B$ mesons is shown in Figure~\ref{fig:fei:mbc-fit} for a signal probability output of the FEI $\mathcal{P}_\mathrm{FEI} > 0.2$. 
The efficiency, which is defined as the fraction of correctly reconstructed $B$ mesons of all $\Upsilon(4\mathrm{S})$ events and the purity, which is defined as the fraction of correctly reconstructed $B$ mesons of all reconstructed $B$ candidates, for two cuts on the FEI signal probability $\mathcal{P}_\mathrm{FEI} > 0.2$ and $\mathcal{P}_\mathrm{FEI} > 0.01$ is tabulated in Table~\ref{tab:fei:eff-pur}.

\begin{table}
	\centering
	\caption{Efficiencies and purities of the FEI for two different signal probability cuts on the FEI output.}
	\label{tab:fei:eff-pur}
	\begin{tabular}{ccrr}
		\hline
		& Candidates & Efficiency & Purity \\
		\hline
		& \multicolumn{3}{c}{FEI Signal Probability $\mathcal{P} > 0.01$}\\
		\hline
		Charged Candidates & 937$\pm$126 & 0.17\% & 24\% \\
		Neutral Candidates & 394$\pm$\hphantom{0}59 & 0.09\% & 25\% \\
		\hline
		& \multicolumn{3}{c}{FEI Signal Probability $\mathcal{P} > 0.2$}\\
		\hline
		Charged Candidates & 389$\pm$\hphantom{0}43 & 0.07\% & 63\% \\
		Neutral Candidates & 182$\pm$\hphantom{0}24 & 0.03\% & 73\% \\
		\hline
	\end{tabular}
\end{table}

\section{Analysis of Inclusive Semileptonic \boldmath{$B \rightarrow X\ell\nu$} Decays}
We performed the analysis of inclusive $B \rightarrow X e \nu$ decays  on the Belle~II data sample of $500\, \mathrm{fb}^{-1}$ from the commissioning run, which is motivated by a similar analysis by the CLEO collaboration~\cite{cleo}.
Due to the lack of an off-resonance data sample only the rediscovery of the decay can be achieved. Statements about $V_\mathrm{ub}$, $V_\mathrm{cb}$ and branching fractions are not possible, because no background subtraction can be performed.

\begin{figure}
	\includegraphics[width=80mm]{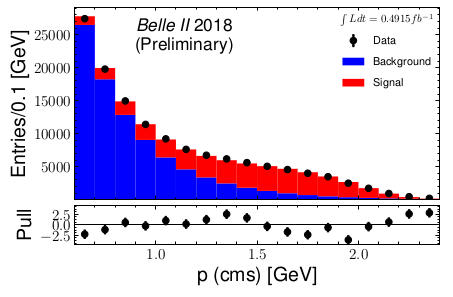}
	\caption{Results of binned maximum likelihood template fits to the data using one component for the background pdf.}
	\label{fig:BXenu:result}
\end{figure}

We select events by requiring one charged electron track originating from the interaction region $|\mathrm{d}z| < 2\, \mathrm{cm}$, $d_0 < 0.5\, \mathrm{cm}$ with at least one hit in the CDC, a center-of-mass momentum of $0.6\, \mathrm{GeV} < p* < 3.3\, \mathrm{GeV}$, a polar angle of the corresponding ECL cluster of $0.767\, \mathrm{rad} < \theta_\mathrm{ECLCluster} < 2.042\, \mathrm{rad}$ which corresponds to the barrel region of the ECL, and a pseudo electron ID of $E_\mathrm{ECL} / p_\mathrm{track} > 0.9$. Background from non-resonant $e^+e^- \rightarrow q\bar{q}$ events is suppressed with the reduced Fox Wolfram moment $R_2 < 0.4$. Contamination of $J/\Psi \rightarrow e^+e^-$ decays is avoided by discarding evens containing a reconstructed $J/\Psi$ candidate with an invariant mass of $3\, \mathrm{GeV} < m_{J/\Psi} < 3.14\, \mathrm{GeV}$.
The distribution of the center-of-mass momentum of the lepton after the event selection is shown in Figure~\ref{fig:BXenu:result}.
The fitted signal yield is $42181\pm304$ over a background yield of $89802\pm374$.

\section{Analysis of Exclusive Semileptonic	\boldmath{$B \rightarrow D^*\ell\mathcal\nu$} Decays}
The decay $B \rightarrow D^*e\nu$ has a large branching ratio of $\mathcal{B}(B \rightarrow D^*e\nu)=(5.05\pm0.14)\,\%$ making them an excellent candidate for a rediscovery with a data sample of $366\, \mathrm{pb}^{-1}$ from the commissioning run.

Events are selected as follows. The impact parameters of the $\pi$ and $K$ tracks are required to be close to the interaction region: $d_0 < 0.5\, \mathrm{cm}$, $|z_0| < 3.0\, \mathrm{cm}$. The center-of-mass momentum of the slow $\pi$ from the $D^\star \rightarrow D$ transition is selected as $p^*(\pi) < 0.4\, \mathrm{GeV}$. The allowed invariant mass range for $D$ candidates is $1.85\, \mathrm{GeV} < m(D) < 1.88\, \mathrm{GeV}$. The allowed mass difference between the $D$ and $D^*$ candidates is $0.144\, \mathrm{GeV} < \Delta m < 0.148\, \mathrm{GeV}$. 
Lepton tracks are required to have impact parameters of $d_0 < 2\, \mathrm{cm}$ and $|z_0| < 5\, \mathrm{cm}$ and a center-of-mass momentum of $1.2\, \mathrm{GeV} < p^*_l < 2.4\, \mathrm{GeV}$. Further, a pseudo electron ID of $E_\mathrm{ECL} / p_\mathrm{track} > 0.94$ is required. Background from non-resonant $e^+e^- \rightarrow q\bar{q}$ events is suppressed with the reduced Fox Wolfram moment $R_2 < 0.25$.

\begin{figure}
	\includegraphics[width=80mm]{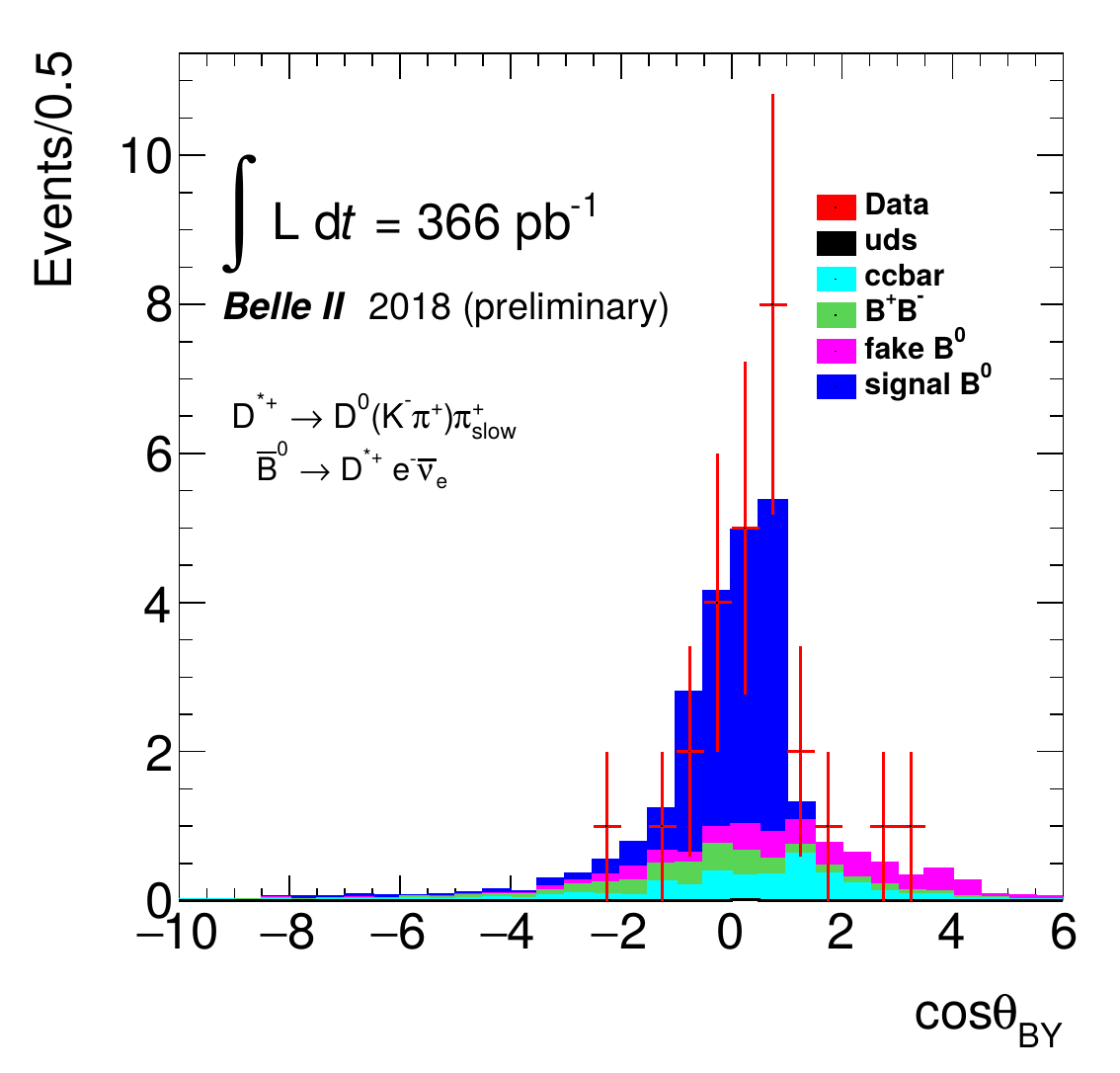}
	\caption{The $\cos \Theta_\mathrm{BY}$ distribution for $B \rightarrow D^*e\nu$ candidates in the data (points with error bars) overlaid with the combination of MC events. All cuts are applied. The MC histograms are scaled to the number of data events recorded.}
	\label{fig:BDstarenu:result}
\end{figure}

The signal extraction is performed on the variable 
\begin{equation}
	\cos \Theta_\mathrm{BY} = \frac{2E_B^* E_Y^* - M_B^2 - m_Y^2}{2p_B^*p_Y^*} \,
\end{equation}
where $E_Y^*$, $p_Y^*$ and $m_Y$ are the center-of-mass energy, center-of-mass momentum and the invariant mass of the $D^*l$ system. Further $M_B$ denotes the nominal $B$ mass, and $E_B^*$ and $p_B^*$ the center-of-mass energy and momentum inferred from the beam energies. For correctly reconstructed candidates, ignoring detector resolution and mis-reconstructed events, this variable peaks in $\cos \Theta_\mathrm{BY} \in [-1, 1]$. The distribution of this variable after the event selection is shown in Figure~\ref{fig:BDstarenu:result}.
We observe a total of 22 events, thereof 15 events in the signal region $\cos \Theta_\mathrm{BY} \in [-1, 1]$ where we expect from MC that 13 events are signal.

\section{Future Prospects of Belle~II}
Belle~II has a rich physics program~\cite{b2pb} which will be pursued in the future. Here two long standing tensions with the SM and how Belle~II will be able to resolve them are shown.

\begin{figure}
	\includegraphics[width=80mm]{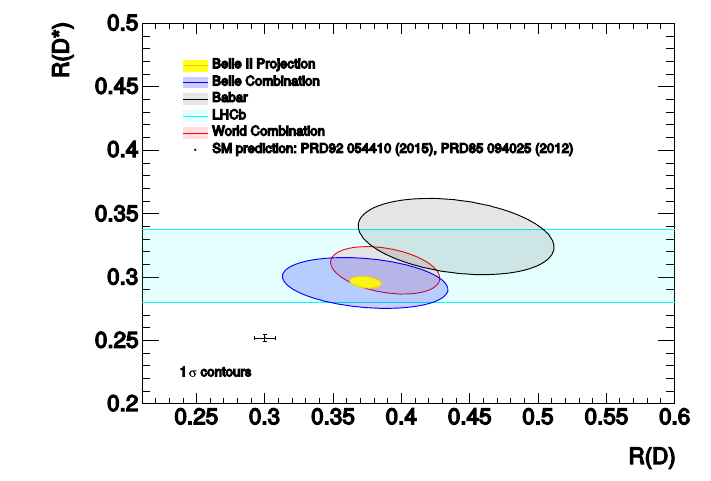}
	\caption{Expected Belle II constraints on the $R(D)$ vs.\, $R(D^*)$ plane compared to existing experimental constraints from Belle. The SM prediction is indicated by the black point with theoretical error bars. The figure is taken from Reference~\cite{b2pb} and does not include the latest Belle measurement from Reference~\cite{RDupdate}.}
	\label{fig:prospects:RDDstar}
\end{figure}

The decays $B \rightarrow D^{(*)}\tau\nu$ are described at the quark level as $b \rightarrow c\tau\nu$ tree-level transitions that proceed in the SM through the exchange of a virtual W boson. The ratios, defined as
\begin{equation}
	R(D^{(*)}) = \frac{\mathcal{B}(B \rightarrow D^{(*)}\tau\nu)}{\mathcal{B}(B \rightarrow D^{(*)}l\nu)} \,
\end{equation}
with $l=e, \mu$, are excellent probes for new physics as theoretical uncertainties from form factors and the CKM matrix element $|V_\mathrm{cb}|$ cancel. The observable has shown large tensions in the past and Belle~II will be able to give a final answer if new physics, e.g. charged Higgs bosons~\cite{np1, np2} or Leptoquarks~\cite{np3}, contribute to the process. The projection of the Belle~II sensitivity in the $R(D)$-$R(D^*)$ plane is shown in Figure~\ref{fig:prospects:RDDstar}.

The determination of the CKM matrix element $|V_\mathrm{ub}|$ from semileptonic $B\rightarrow X_\mathrm{u}l\nu$ decays exhibits a long-standing tension to the determination of $|V_\mathrm{ub}|$ from exclusive $B\rightarrow\pi l\nu$ decays and global fits of the CKM unitarity triangle. With the full Belle~II data sample the decay $B\rightarrow \tau \nu$ will become a competitive method to extract $|V_\mathrm{ub}|$. The current experimental status and projections for $5\, \mathrm{ab}^{-1}$ and $50\, \mathrm{ab}^{-1}$ are shown in Table~\ref{tab:prospects:Vub}. 

\begin{table*}
	\caption{Expected errors in $|V_\mathrm{ub}|$ measurements with the Belle full data sample, $5\, \mathrm{ab}^{-1}$ and $50\, \mathrm{ab}^{-1}$ Belle II data. Note that the statistical error quoted for exclusive $|V_\mathrm{ub}|$ branching fraction, however a fit to the spectrum information is used to determine $|V_\mathrm{ub}|$. We use the		lattice-QCD projected precision for the future data sets. From Ref.~\cite{b2pb}.}
	\label{tab:prospects:Vub}
	\begin{tabular}{cccccc}
		\hline
		& Statistical & Systematic & Total Exp. & Theory & Total \\
		& & (reducible, irreducible) & & & \\
		\hline
		$|V_\mathrm{ub}|$ exclusive (had. tagged) & & & & & \\
		$711\, \mathrm{fb}^{-1}$ & 3.0 & (2.3, 1.0) & 3.8 & 7.0 & 8.0 \\
		$5\, \mathrm{ab}^{-1}$ & 1.1 & (0.9, 1.0) & 1.8 & 1.7 & 3.2 \\
		$50\, \mathrm{ab}^{-1}$ & 0.4 & (0.3, 1.0) & 1.2 & 0.9 & 1.7 \\
		\hline
		$|V_\mathrm{ub}|$ exclusive (untagged) & & & & & \\
		$605\, \mathrm{fb}^{-1}$ & 1.4 & (2.1, 0.8) & 2.7 & 7.0 & 7.5 \\
		$5\, \mathrm{ab}^{-1}$ & 1.0 & (0.8, 0.8) & 1.2 & 1.7 & 2.1 \\
		$50\, \mathrm{ab}^{-1}$ & 0.3 & (0.3, 0.8) & 0.9 & 0.9 & 1.3 \\
		\hline
		$|V_\mathrm{ub}|$ inclusive & & & & & \\
		$605\, \mathrm{fb}^{-1}$ (old B tag) & 4.5 & (3.7, 1.6) & 6.0 & 2.5−4.5 & 6.5−7.5 \\
		$5\, \mathrm{ab}^{-1}$ & 1.1 & (1.3, 1.6) & 2.3 & 2.5−4.5 & 3.4−5.1 \\
		$50\, \mathrm{ab}^{-1}$ &0.4 & (0.4, 1.6) & 1.7 & 2.5−4.5 & 3.0−4.8 \\
		\hline
		$|V_\mathrm{ub}|$ $B\rightarrow\tau\nu$ (had. tagged) & & & & & \\
		$711\, \mathrm{fb}^{-1}$ & 18.0 & (7.1, 2.2) & 19.5 & 2.5 & 19.6 \\
		$5\, \mathrm{ab}^{-1}$ & 6.5 & (2.7, 2.2) & 7.3 & 1.5 & 7.5 \\
		$50\, \mathrm{ab}^{-1}$ & 2.1 & (0.8, 2.2) & 3.1 & 1.0 & 3.2 \\
		\hline
		$|V_\mathrm{ub}|$ $B\rightarrow\tau\nu$ (SL tagged) & & & & & \\
		$711\, \mathrm{fb}^{-1}$ & 11.3 & (10.4, 1.9) & 15.4 & 2.5 & 15.6 \\
		$5\, \mathrm{ab}^{-1}$ & 4.2 & (4.4, 1.9) & 6.1 & 1.5 & 6.3 \\
		$50\, \mathrm{ab}^{-1}$ & 1.3 & (2.3, 1.9) & 2.6 & 1.0 & 2.8 \\
		\hline
	\end{tabular}
\end{table*}

\section{Conclusion}
With the first $500\, \mathrm{pb}^{-1}$ recorded by Belle~II during the 2018 commissioning run, we have successfully applied the newly developed algorithms and tools on real data. The FEI shows excellent performance, and we observe clear signals for inclusive $B\rightarrow Xe\nu$ and exclusive $B\rightarrow D^* e \nu$ decays. With the target integrated luminosity of $50\, \mathrm{ab}^{-1}$ Belle~II will be able to give a final answer on long standing tensions with the SM.

\bigskip 

\end{document}